\documentclass{aa}
\usepackage{graphicx}                    
\usepackage{txfonts}

\begin{document}

\title{Polar cap magnetic field reversals during solar grand minima: could pores play a role?}

\author{Michal \v{S}vanda\inst{1,2}
\and
Allan Sacha Brun\inst{3}
\and
Thierry Roudier\inst{4}
\and
Laur\`ene Jouve\inst{4,3}
}
\institute{Astronomical Institute (v. v. i.), Czech Academy of Sciences, Fri\v{c}ova 298, CZ-25165 Ond\v{r}ejov, Czech Republic
\email{michal@astronomie.cz}
\and
Astronomical Institute, Charles University in Prague, Faculty of Mathematics and Physics, V Hole\v{s}ovi\v{c}k\'ach 2, CZ-18000 Prague 8, Czech Republic
\and
Laboratoire AIM Paris-Saclay, CEA/Irfu Universit\'e Paris-Diderot CNRS/INSU, F-91191 Gif-sur-Yvette, France
\and
Institut de Recherche en Astrophysique et Plan\'etologie, Universit\'e de Toulouse, CNRS, 14 avenue \'Edouard Belin, F-31400 Toulouse, France}
\abstract{
We study the magnetic flux carried by pores located outside active regions with sunspots and investigate their possible contribution to the reversal of the global magnetic field of the Sun. We find that they contain a total flux of comparable amplitude to the total magnetic flux contained in polar caps. The pores located at distances of 40--100~Mm from the closest active region have systematically the correct sign to contribute to the polar cap reversal. These pores can predominantly be found in bipolar magnetic regions. We propose that during grand minima of solar activity, such a systematic polarity trend, akin to a weak magnetic (Babcock-Leighton-like) source term could still be operating but was missed by the contemporary observers due to the limited resolving power of their telescopes.}

\maketitle 

\section{Pores as proxies of Babcock-Leighton source terms}

The Sun is a magnetically active star that exhibits a large range of dynamical phenomena at its surface and in its atmosphere directly related to this activity. The magnetically related activity has been directly observed for more than four centuries and such record clearly demonstrates that it has been modulated over time variously in its amplitude, frequency of the occurrence of the active phenomena and the location of their occurrence. In a 11-yr cycle, the global sign of solar magnetic field reverses. The reversal of the global magnetic field is prominent in the polar regions -- caps -- and occurs usually close to the maximum of the cycle \citep{2004AA...428L...5B}. In the current paradigm, fields originating from the decay of active regions (mostly from their trailing parts) are transported towards the poles by the meridional flow. It is believed that these organised diffuse fields are effectively responsible for the polar field reversals not only at the surface, but possibly deep within the convective envelope leading to a global reversal. 

The solar magnetic field is thought to be due to a physical regeneration mechanism termed fluid dynamo, which converts mechanical energy contained in convective motions into magnetic energy. How such conversion processes operate is subject to debate but it is generally thought that a large-scale shear regenerates the toroidal component of the field whereas the poloidal component is created via either helical turbulent motions, also known as the $\alpha$-effect, or through the decay of tilted active regions \citep{1961ApJ...133..572B,leighton69,1999ApJ...518..508D,2013SSRv..tmp..100B,2015Sci...347.1333C}. This latter process, usually called the Babcock-Leighton (BL) mechanism, entirely rests on the presence of tilted active regions that continuously emerge at the surface of the Sun during the 11-yr activity cycle. When the magnetic field is strong enough, sunspots may form in these active regions \citep{1994ApJ...433..867P}. 

There are however periods of very low activity in the recorded activity indices, when the appearance of spots was rather rare according to the observations. These periods \citep[termed ``deep'' or ``grand'' minima -- ][]{1993AA...276..549R,2002AA...395.1013B,2013LRSP...10....1U} are a challenge for the BL dynamo models, as the necessary surface term apparently vanishes. During these large minima observers were still staring at the Sun, however they recorded very few sunspots \cite[e.g. during the Maunder minimum between 1645--1715, see][]{1976Sci...192.1189E,1996SoPh..165..181H}. Although there was a general conclusion that the solar activity was lower than normal, solar proxies (such as $^{10}$Be isotope concentration) suggest that the 11yr cycle was weak but fairly regular during the Maunder minimum \citep{1998SoPh..181..237B}. Similarly, the geomagnetic activity showed a clear 11yr variation \citep{1998GeoRL..25..897C}. 

When taking the resolving power of the telescopes during the Maunder minimum era (which is estimated to be 2--5"), it is possible that the BL term was still operational, only the organised magnetic field were weak so that the sunspots could not form regularly. In this particular study, we choose to focus on a certain class of active regions without sunspots: those containing pores. Pores are white-light features similar to small sunspots. They are believed to be strong concentrations of magnetic field and therefore to potentially possess enough flux for the reversal of the polar field. Therefore, pores might be important agents of the organised magnetic field and were possibly invisible to the ancient observers. Compared to sunspots the pores have a simpler configuration of the magnetic field, which is oriented mostly vertically \citep{1964suns.book.....B} with strength between 1700~G \citep{1996AA...316..229K} and 2600~G \citep{1982SoPh...80..251B}, depending on the observed spectral line. The pores do not contain a penumbral structure, the majority of them has a diameter of 1500--3500~km, however some are no larger than granules (700--1500~km). Pores larger than 4500~km (that corresponds roughly to 5.5") are uncommon. 

Unlike sunspots that are mainly confined to a relatively narrow belt ($\pm35\degr$) around the solar equator, the pores were observed up to 75$\degr${} \citep{1955QB521.W3.......}. Also magnetic flux concentrations are spread all the way to solar poles. \cite{2012ApJ...753..157S} analysed high-resolution Hinode data and found a variety of flux concentrations, some of them having a magnetic flux comparable to the flux of pores. However, no pores in polar regions were detected in the visible-light images. 

\section{Purpose of the present work and methodology}

In the present work, the idea is to address the following question: if we discard the magnetic field in active regions with sunspots, will there still be enough flux to cause the reversal of the polar cap? As said before, we will focus on regions containing pores, as they are believed to be weaker than spotted active regions (SARs henceforth) but still contain a significant amount of flux. Moreover, we assume here that such regions with smaller activity were present on the Sun also during grand minima, as the geomagnetic and heliospheric activity indices tend to indicate. \cite{1998A&A...333.1053L} showed that during the evolution of an active region the pores start to appear soon after the emergence of the magnetic field into the photosphere. However, the standard scenario of an evolution of the active region \citep[as described by e.g.][]{2015LRSP...12....1V} may stop before the formation of proper sunspots. These regions form a low-end tail of the size- and magnetic flux-spectrum of active regions \citep[e.g.][]{2003ApJ...584.1107H}. Therefore pores appear both near the sunspots and in isolation \citep{1999ApJ...510..422K}, even though it seems that in the space of fundamental parameters, pores and sunspots depict distinct groups \citep{2015ApJ...811...49C} so one cannot simply say that a pore is a ``small sunspot''. Yet, it is believed that pores are manifestations of rising flux tubes in the same way as sunspots are. 

To study the ability of pores lying outside SARs to maintain the reversal of global polarity of the Sun even during grand minima, we thus have to go through several steps:

\begin{itemize}
\item Identify the pores lying outside active regions possessing spots for cycles 23 and 24 where SOHO/MDI and SDO/HMI were and still are available. We consider all active regions containing sunspots and/or pores, however we discard all sunspots and also pores lying close to the ``cores'' of spotty active regions. Examples may be seen in Fig.~\ref{fig:scheme}, where the ellipse centred at roughly $(1800,2300)$ encircles the region with both sunspots and pores, where the core was discarded, whereas the ellipse centred roughly at $(2150,2500)$ indicates an active region containing only pores. The third remaining active region in this example contains both pores and sunspots only in the ``core'' and all these magnetic features are discarded from the selection.

\item For those pores, identify a possible polarity bias. More specifically, we would like to know if those pores show a polarity dominance in each hemisphere, and if this dominant polarity is indeed opposite to the polar cap in the rising phase of the cycle and of the same sign as the polar cap in the declining phase. Moreover, we give estimates of the flux contained in those detected pores to assess their potential to reverse the polar caps. Indeed, we can only determine their "potential for reversal" since we should keep in mind that this flux will not entirely be transported to the poles. It has to be noted that in a normal cycle, only a fraction of the total magnetic flux in the typical active regions with sunspots (which is in the order of $10^{23}$~Mx) is transported towards the poles \citep[e.g.][]{2014arXiv1410.8867S}.

\item If a polarity bias is indeed identified in the previous step, we have to ensure that this is not influenced by the stronger neighbouring SARs. To do so, we wish to determine the "distance of influence" of those regions by determining the possible relationship between the polarity of our detected pores and the  polarity of the closest SAR. We will then be able to establish to which distance a SAR may influence the polarity of neighbouring weaker active regions. 

\item Finally, clearly identify the pores, if any, with the right polarity bias in each hemisphere, which are not influenced by neighbouring SARs and possess the right amount of flux to potentially reverse the polar caps. Then conclude on their ability to maintain a Babcock-Leighton surface mechanism even during grand minima episodes. 

\end{itemize}



\begin{figure*}[t]
\centering
\includegraphics[width=0.7\textwidth]{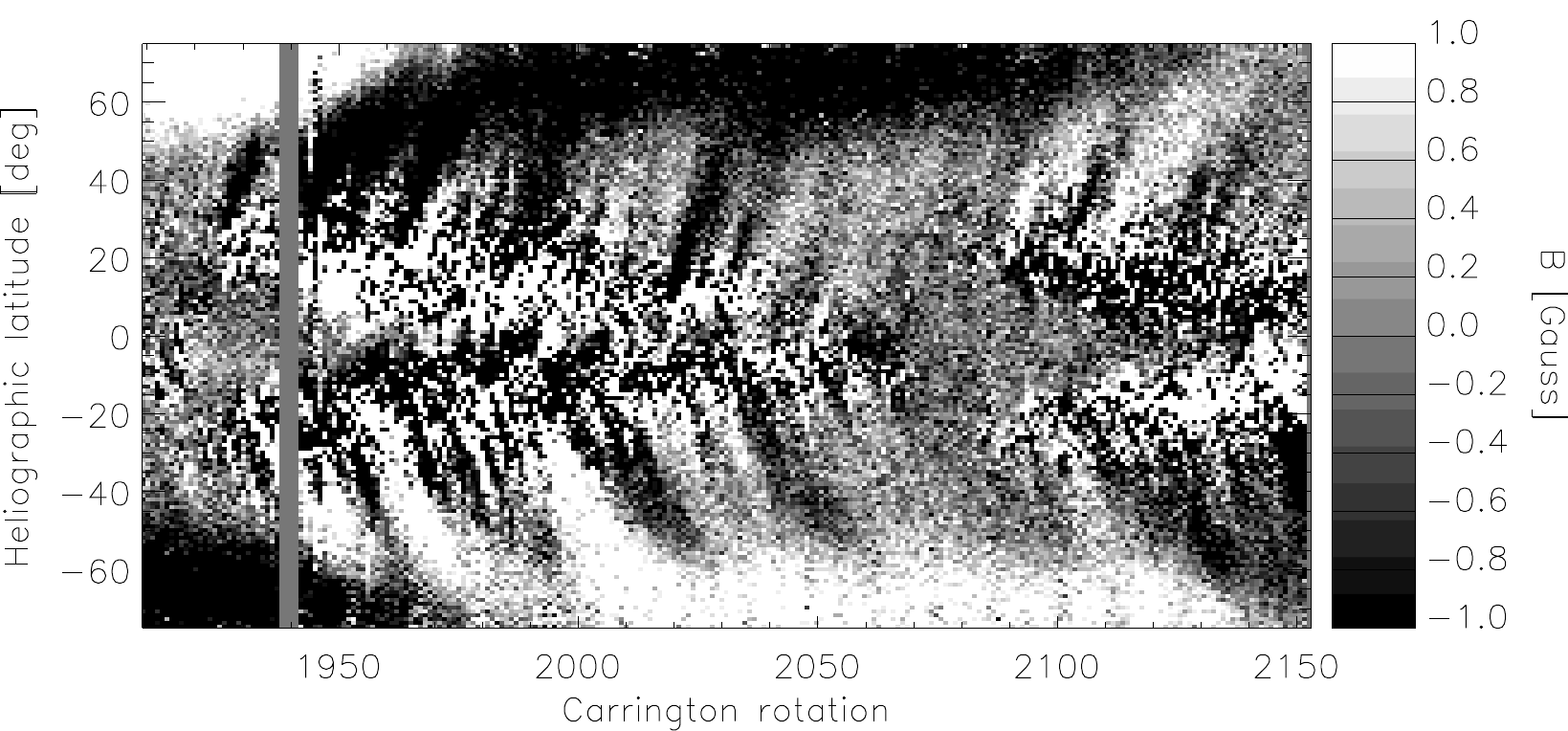}\\
\includegraphics[width=0.7\textwidth]{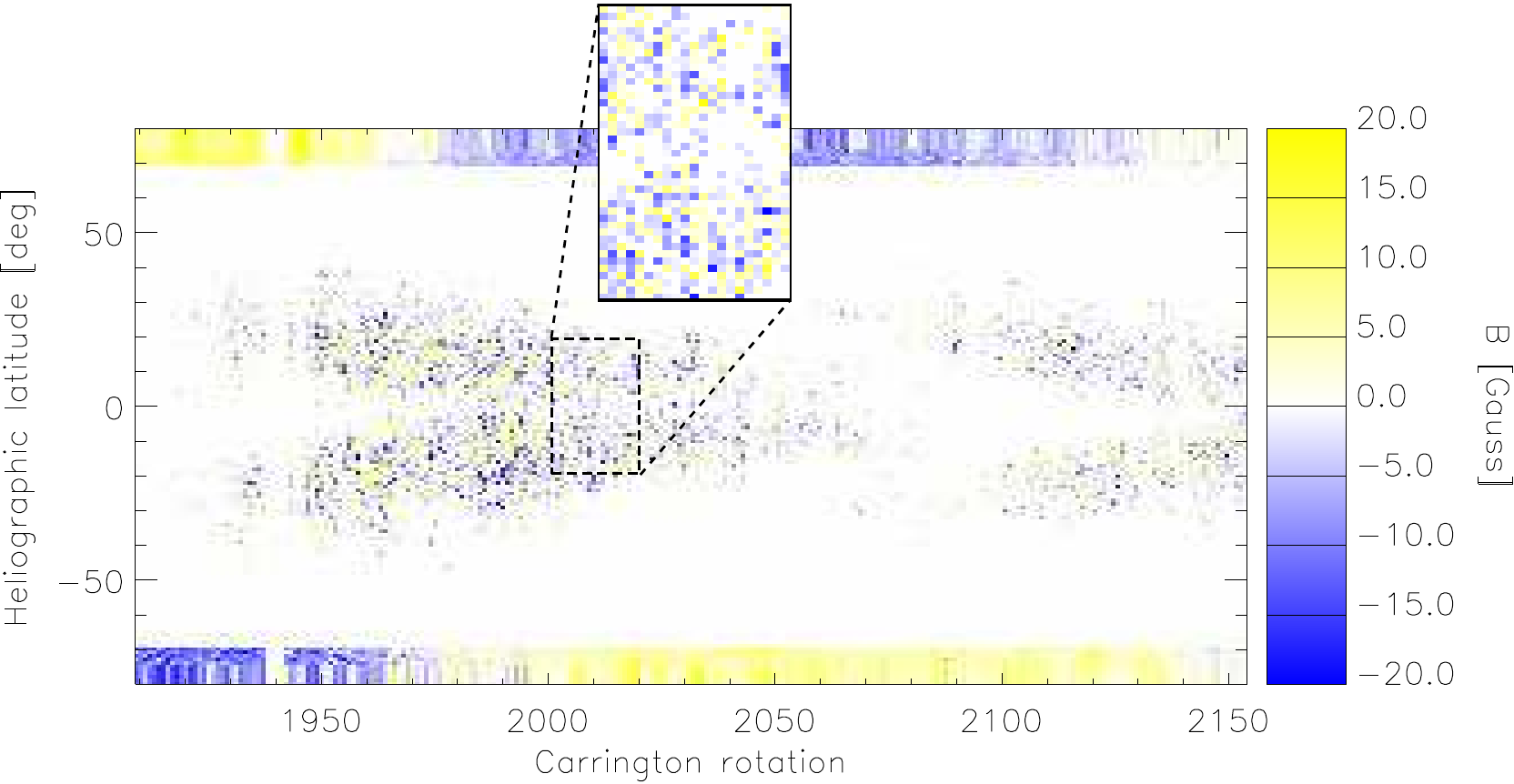}
\caption{Top -- Magnetic butterfly diagram from MDI and HMI synoptic maps shows the prevailing polarity of the magnetic field in polar caps, reversals of the global magnetic field and the flux transported to the polar regions from the relics of the active regions. Bottom -- Magnetic butterfly diagram of the pores outside active regions. At the top and the bottom the polar cap from magnetic butterfly from above is inserted. The intensity of the average magnetic field in the inserts is boosted by the factor of 5 to make them visible. This figure clearly shows that the pores follow the cycle. The inset in the lower panel demonstrates the mixed polarity in magnification.}
\label{fig:mgbtfly}
\end{figure*}

\section{Data and Methods}
\label{sect:data}

\subsection{Data}
We used strictly the archives of measurements obtained by space-borne synoptic experiments. To cover the whole cycle 23 and a progressed part of cycle 24, we combined the datasets from SOHO/MDI\footnote{We used full-disc intensitygrams downloaded from the online archive at sohowww.nascom.nasa.gov/data/archive/index\_ssa.html (there are at most 4 images per day) and corresponding magnetograms stored in the series {\tt mdi.fd\_m\_96m\_lev182} at jsoc.stanford.edu} (covering 19 May 1996 to 12 January 2010) and SDO/HMI\footnote{Full-disc intensitygrams and magnetograms were taken directly from the series {\tt hmi.ic\_45s} and {\tt hmi.m\_45s} at jsoc.stanford.edu with hourly cadence} (covering 29 March 2010 to 24 August 2014). During the processing the HMI data were mimicked to be MDI-like (see Appendix~\ref{sect:appendixa} for details), hence effectively the sampling of 6~hours was used. Higher cadence was used for testing purposes only.

\begin{figure*}[!t]
\centering
\includegraphics[width=\textwidth]{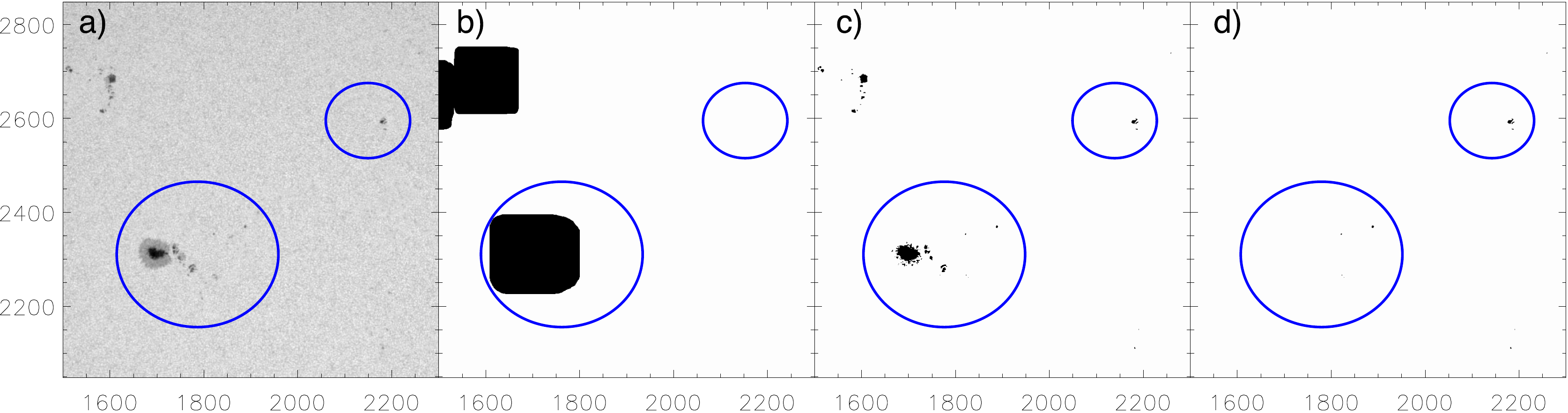}
\caption{Steps in the detection of pores outside SARs. \emph{a} -- part of the original HMI intensitygram on 4 Oct 2014 (axes indicate the pixel positions in the original image). \emph{b} -- mask indicating the positions of active regions. \emph{c} -- mask indicating the detection of all dark features (i.e. spots and pores). \emph{d} -- the final mask segmenting only pores outside SARs. The figures displays a section of the full-disc frame captured on 4 October 2014 at 01:00:00 TAI. The ticks on both axes are in pixels, the disc centre is located at coordinates $(2042,2048)$. The pixel size corresponds to 0.5".}
\label{fig:scheme}
\end{figure*}
\subsection{Supplementary material}
To assess the total magnetic flux in the polar caps together with the assessment of the prevailing magnetic polarity in a given solar hemisphere at a given time, we constructed the magnetic butterfly diagram (the map saturated at $\pm1$~Gauss is shown in Fig.~\ref{fig:mgbtfly}, top panel). Such diagram was constructed from the Carrington-rotation synoptic maps\footnote{These maps are available for MDI era at soi.stanford.edu/magnetic/index6.html (we used Carrington rotations 1909-2104), and from series {\tt hmi.Synoptic\_Ml\_720s} at jsoc.stanford.edu for HMI era (Carrington rotations 2105--2153)}. The synoptic maps were averaged over the longitude into a column vector, which forms one column of the resulting magnetic-butterfly map. Since the diagram is computed by means of averaging, the units are in G=Mx$\cdot$cm$^{-2}$. If one needs to get a total net magnetic flux (in Mx), one has to multiply this number by an appropriate surface area.

\subsection{Pores detection}
\label{sect:detection_procedure}
We used full-disc intensitygrams and an automatic IDL procedure to search for the pores. The routine is based on image segmentation and searches for regions of a given size significantly darker than the surroundings (by means of thresholding). The search is performed in two steps, differing by the size of the structures that are being looked for. In the first step, rather large areas, corresponding to  active regions with sunspots, are searched for. The mask is obtained by dilation of detected features by 100 pixels, which will be used to exclude the pores in SARs. The small structures, corresponding to the pores, are then searched for in the second step. In this step, also artifacts such as dust or bad pixels are found. These artifacts are removed following the procedure described in Appendix~\ref{sect:appendixa}. 

From the set of detected small features only those outside the mask of active regions are kept for further analysis. These features are nicknamed as \emph{pores} henceforth. All steps are demonstrated on an example in Fig.~\ref{fig:scheme}. The segmentation and labelling is done using standard IDL routines. For each pore, the total area $S$ in squared pixels is computed (by taking into account the projection effect, hence dividing the projected area by a cosine of the heliocentric angle), the mean magnetic field intensity (the location of each pore is used as a mask for the corresponding line-of-sight magnetogram, hence the l.o.s. magnetic field is known within the pore), and the positions of the pore, in both CCD coordinates (in pixels) and Carrington coordinates (in degrees). Some other useful quantities are also stored, such as the heliocentric angle or effective radius ($r_{\rm eff}=\sqrt{S/\pi}$). 
 
\section{General polarity trends and fluxes for pores detected in cycles 23 and 24}
\label{sect:appearance}
The appearance of the pores in terms of their positions as a function of time was studied from the combined MDI+HMI data set. All pores fulfilling requirements described in Appendix~\ref{sect:appendixa} were taken into account. The pores were placed into large maps of various physical quantities describing the pores (average magnetic flux in Gauss, measured area, number of pores in the given coordinate bin) in Carrington coordinates, with latitudinal extent $\pm80$~degrees and a continuously increasing longitude from 0 (beginning of CR1909) to 88\,560 degrees (end of CR2153). The map was formed with a binsize of 1 degree on both axes. 

First, we constructed an equivalent of the magnetic butterfly diagram, the line-of-sight magnetic field intensity in pores as a function of time and latitude. Similarly to the construction of the reference magnetic butterfly diagram from synoptic maps we averaged the magnetic flux in pores over longitude separately for each Carrington rotation. The magnetic butterfly diagram of pores outside SARs is plotted in Fig.~\ref{fig:mgbtfly}, bottom panel. From this figure we see that the pores located outside SARs follow the cycle migration for their position, contributing to broadening the activity belt. From the first sight, however, their polarity seems mixed. It is the same situation as with sunspots. They also appear with a mixed polarity (a bipolar sunspot group has both positive and negative parts) and only after the averaging the dominance of one polarity in the given hemisphere appears. 
\begin{figure}[h]
\centering
\includegraphics[width=0.49\textwidth]{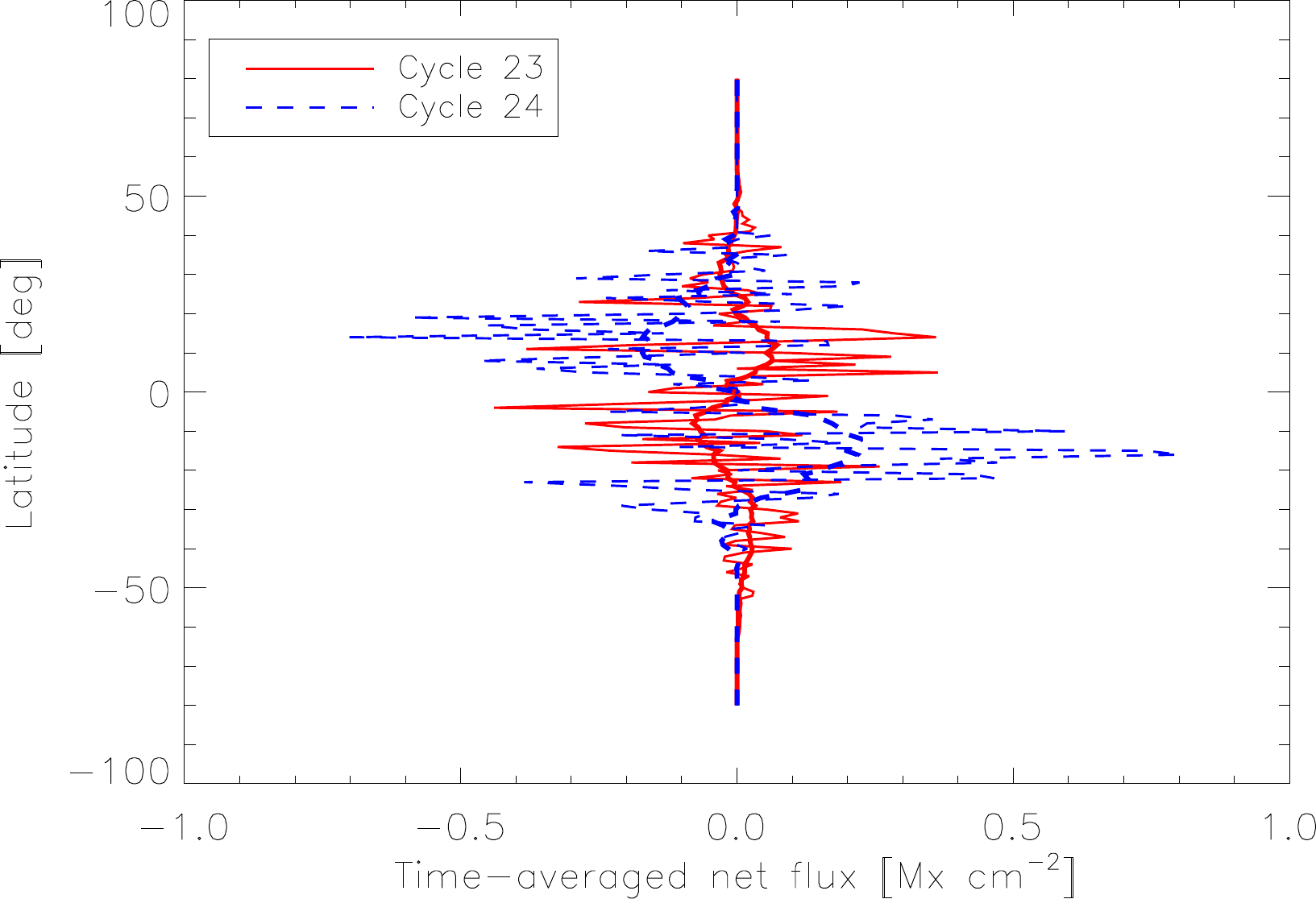}\\
\includegraphics[width=0.49\textwidth]{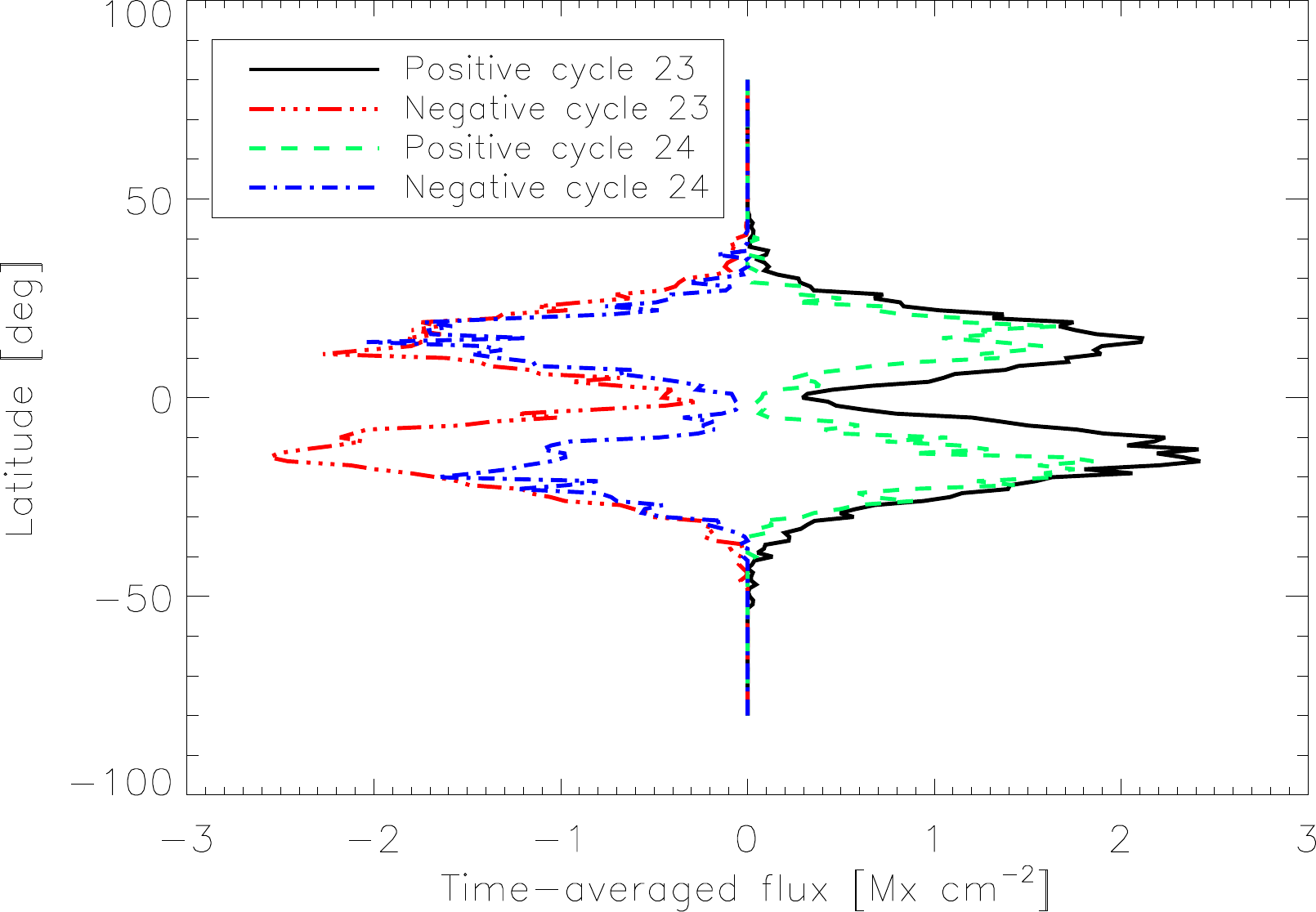}
\caption{Upper panel -- Net magnetic flux in the pores averaged within the cycle over time. In this case the random-like mixed polarity in the pores average out and the secular polarity trends remain. Obviously, in cycle 23 the positive polarity prevails on the northern hemisphere, which reverses in cycle 24. The thick lines represent the smoothed curves. Bottom panel -- Magnetic fluxes stored in pores outside SARs averaged over cycles separately for positive and negative polarities.}
\label{fig:hemispheric_flux}
\end{figure}

To demonstrate the polarity trends in these pores we must filter out the random polarity appearance so that only the polarity bias remains, if present. We averaged the magnetic flux in the pores within each of the investigated cycles. The plot is displayed in Fig.~\ref{fig:hemispheric_flux} in the upper panel. From looking at the envelope of the noisy curve there seems to be a dominance of the positive flux in the northern hemisphere and negative flux in the southern hemisphere in cycle 23, whereas the sign reverses for both hemispheres in cycle 24.

Using such plot we may roughly estimate the total magnetic flux in the pores by multiplication of the average magnetic field intensity by the corresponding area. The average magnetic field intensity (estimated to be 0.1~G) was computed using the latitudinal band between $0\degr${} and $60\degr${}, which roughly corresponds to an area of $2.7\times10^{22}$~cm$^2$. That would roughly correspond to a total net magnetic flux of $2.7\times10^{21}$~Mx. For comparison, the total magnetic flux contained in the polar cap may be similarly estimated from the magnetic butterfly diagram (Fig.~\ref{fig:mgbtfly}), where the average magnetic field intensity may be estimated to be 1~G\footnote{Such value is consistent with the measurements of the polar fields performed by Wilcox Solar Observatory and published at http://wso.stanford.edu/gifs/Polar.gif} and the area of the polar cap (everything above $70\degr${} of latitude) as $4\times10^{21}$~cm$^2$. The total magnetic flux in the polar cap is estimated to be $4\times10^{21}$~Mx. In the literature, larger total fluxes in polar caps were reported: \cite{1961ApJ...133..572B} $8\times 10^{21}$~Mx (estimated in the minimum between cycles 18 and 19), \cite{1966ApJ...144..723S} quoted varying polar flux between $6\times 10^{21}$ to $2.1\times 10^{22}$~Mx between years 1905--1964 with $1.2\times 10^{22}$~Mx being the typical value of the polar flux at the maximum of activity, and \cite{2004AA...428L...5B} reported $1.5\times 10^{22}$ to $3.7\times 10^{22}$~Mx depending on the data used for period of 1996--2003. A more sophisticated determination of the total polar magnetic flux from various data sets was done by \cite{2012ApJ...753..146M}, where the magnetic fluxes in polar caps were found to be around $10^{22}$~Mx. So it would seem that a value of $10^{22}$~Mx is a representative number of the total magnetic flux in polar caps in solar cycles in the 20th and 21st centuries and half this value may be considered representative for a weak cycle like cycle 24 (as a halved amplitude of polar field intensity compared to cycle 23 suggests). This trend suggests that during very weak cycles, e.g. during the grand minima, the total polar flux may be even smaller. 

Is there a difference in the flux carried by the pores outside SARs in cycle 23 and 24? Additional plot (Fig.~\ref{fig:hemispheric_flux}, the bottom panel) is constructed. That represents averages over cycles 23 and 24 of the butterfly diagram of the pores (Fig.~\ref{fig:mgbtfly} bottom panel) separately for pores with negative and positive magnetic field. In the northern hemisphere, there seems to be a comparable amount of magnetic flux of both polarities in the pores in both cycles. In the south, there is much less magnetic flux in the pores in cycle 24 than in cycle 23. Note that there has also been less sunspots in the South in cycle 24 than there were in cycle 23, at least until 2014 when the southern hemisphere became more active. 

One should keep in mind that the analysis of cycle 24 is incomplete, as roughly half of this cycle progressed so far, hence the analysis might be biased. The routines resolved 21\,403 pores over cycle 23 and 3\,969 over an incomplete (roughly half) cycle 24. Taking into account this possible bias, we can still conclude that the pores in cycle 24 have on average a larger line-of-sight magnetic field intensity than pores in cycle 23. 

We do not detect any pores at distances larger than $40\degr${} from the equator. \cite{2014AA...563A.112V} mentioned pores located at slightly higher latitudes ($\sim\pm45\degr$), the difference can be attributed to the higher resolution of the Hinode data they used. 

The origin of the polarity trend seen in our pores analysis must be linked to the global organisation of the solar magnetic field through a large scale convective dynamo \citep{2004ApJ...614.1073B,2014arXiv1410.6547A}. Field emergence is believed to be linked to rising subsurface structures whose size, amplitude, twist and fibril state is still debated and the subject of intense theoretical and observational studies \citep[][and references therein]{2008ApJ...676..680F,2013ApJ...762....4J,2013ApJ...762...73N, 2013SoPh..287..215M, 2014SSRv..186..227S}.  In the next two sections we look for spatial and temporal correlation between our selected pores and other magnetic features.

\section{Pores' polarity in cycle 24: influence of neighbouring SARs and relation to the polarity of the polar cap}
\begin{figure*}[!t]
\centering
\includegraphics[width=0.7\textwidth]{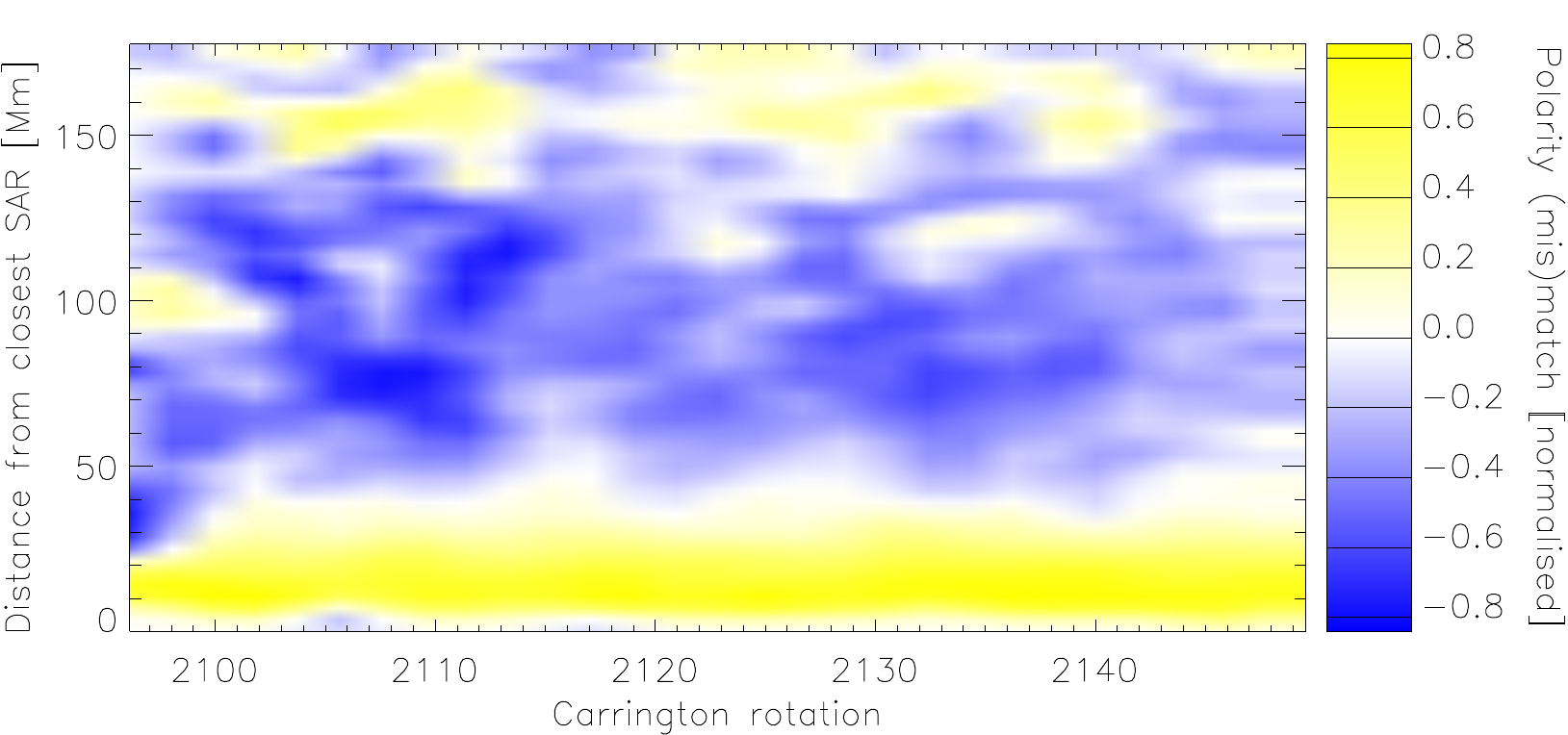}\\
\includegraphics[width=0.7\textwidth]{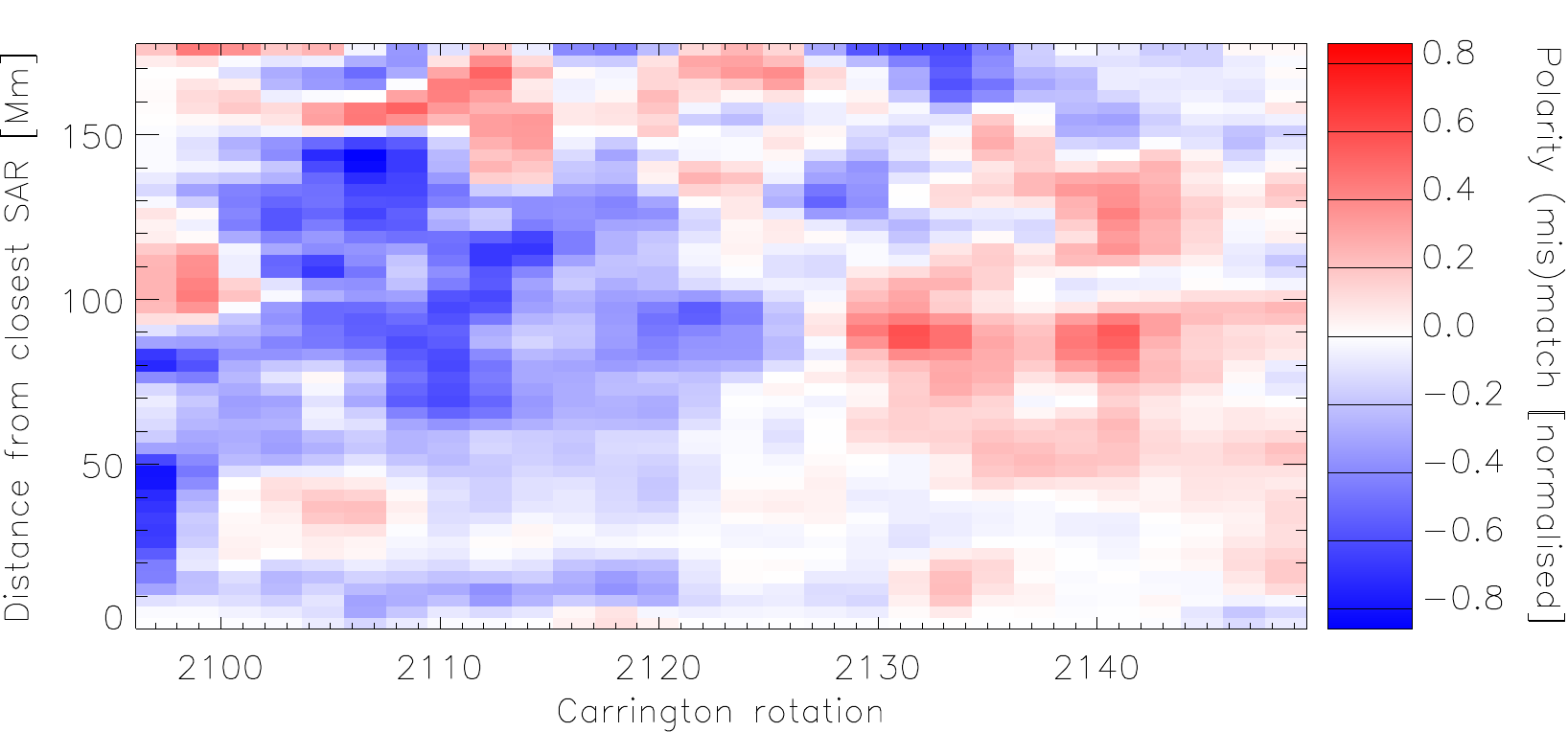}
\caption{Upper panel -- Binned index of match of the flux polarity in pores and their closest SAR as a function of distance from the closest SAR and time. Bottom panel -- Binned index of match of the flux polarity in pores and the adjacent polar cap as a function of distance from the closest SAR and time. One can see that in the intermediate distances 40--100~Mm the polarity systematically does not match that of the corresponding polar cap in the early rising phase of the solar cycle 24, while they match near the maximum of cycle 24. Note: both panels were smoothed to increase the visibility of the polarity biases.}
\label{fig:polarity_match}
\end{figure*}
\label{sect:polarity}
Wanting to investigate the feasibility of the pores outside SARs to contribute to the global field reversals, we studied the possible trends of the pores' polarity with regards to the closest SAR, and also the polarity of the polar cap of the same hemisphere. To investigate this, we used the HMI data only and dropped some of the requirements of the previous analysis. Mainly, we dropped the limiting requirement on the minimal area of the pore. Now we use all detected pores, including those having an area of only 1 px$^2$. For this reason, the total number of pores is much larger. We proceeded in the following way:

\begin{enumerate}
\item For each measurement (frames sampled with a 6-hr cadence from the period from 8 April 2010 to 30 May 2014) the SARs were segmented out using a mask. Similar mask to that of step one in the method outlined in Section~\ref{sect:detection_procedure} was used to isolate the SARs. The segmentation does not necessarily select complete SARs (both poles) within one mask, it might be that very large open bipolar SARs are detected as two. Then for each pore detected in the same frame using the complete procedure described in Section~\ref{sect:detection_procedure}, the closest (distance measured on the sphere) SAR was identified. The distance is measured and stored. The histogram of distances shows a peak around 30~Mm with a FWHM around 18~Mm and a long tail towards larger distances. It drops behind 120~Mm and remains almost flat for larger distances. 
\item For the pore and the closest SAR, the signs of the flux were compared. A ``match'' tag of the pore is set to 1 in case the signs agree and to $-1$ in case they do not. The flux within the SAR is taken as an average over the whole mask. Usually the total flux in the leading part is larger than in the trailing part, hence the sign of flux within the SAR used in the comparison is that of the leading part. Should the SAR be large, bipolar in intensity and open, then it may be detected as two. In such case the sign considered in the comparison is that of the closest pole of such SAR. The methodology of the segmentation is such that the latter case occurs only when the poles of the SAR are at least 60~Mm apart with absolutely no spots or pores in between.
\item Similarly, the signs of the flux in the pore and the closest polar cap at the given time, read out from the butterfly diagram, is compared. Again, a ``match'' tag of the pore is set similarly to the previous step. 
\end{enumerate}

\noindent In the ongoing processing, the pores are binned in the distance-to-the-closest-active-region space (with binsize of 5~Mm) and in time (binsize of 54 days) and the corresponding tags are averaged. The sizes of the bins were selected so that the resulting maps are sufficiently smooth, as the filling factor of the pores in the maps is low. Note that the binsize in the temporal domain corresponds roughly to two Carrington rotations. Hence for the given distance and time, if there is a statistically significant match between the polarity of the pore and the polarity of the closest SAR, the averaged-tag value should approach unity. In case there is systematically a mismatch of the two, the averaged-tag value should approach the value of $-1$. In case there is no obvious rule, the averaged-tag value should be around zero. 

In total, more than 118\,000 individual pores were studied. The top panel of Figure~\ref{fig:polarity_match}  demonstrates the relation between the pore and the closest SAR. We see two conclusions: (1) There seems to be a significant match between the polarity of the pore and the closest SAR for the pores located between 10 and 40 Mm from the closest AR, and (2) there seems to be a significant mismatch in the polarity of the pore and the closest AR for the pores located between 40 and 140~Mm from the closest AR.

For the comparison between the pore's polarity and the polarity of the corresponding polar cap (see Fig.~\ref{fig:polarity_match} the bottom panel) the conclusions are different, also with less clear trends. (1) Pores around 0--40~Mm (most of the pores) from the AR seem to have a mixed polarity compared to the polarity of the corresponding cap. (2) Pores in distances 40 to 100~Mm (possibly even to distance of 140~Mm) from the closest AR seem to have the opposite polarity to the cap in the rising phase of the cycle, and the same polarity as the cap in the late rising/plateau phase of the cycle. No conclusions can be made with regards to the declining phase of cycle 24, as it is still progressing. The trends do not change when both northern and southern hemispheres are investigated separately, including the phase shift of the ``reversal'' timing in accordance to the delayed polarity reversal in the southern cap in cycle 24 (by about two years). Pores located farther than 150~Mm from the closest SAR do not seem to depict any systematic behaviour. 

\begin{figure}[h]
\centering
\includegraphics[width=0.49\textwidth]{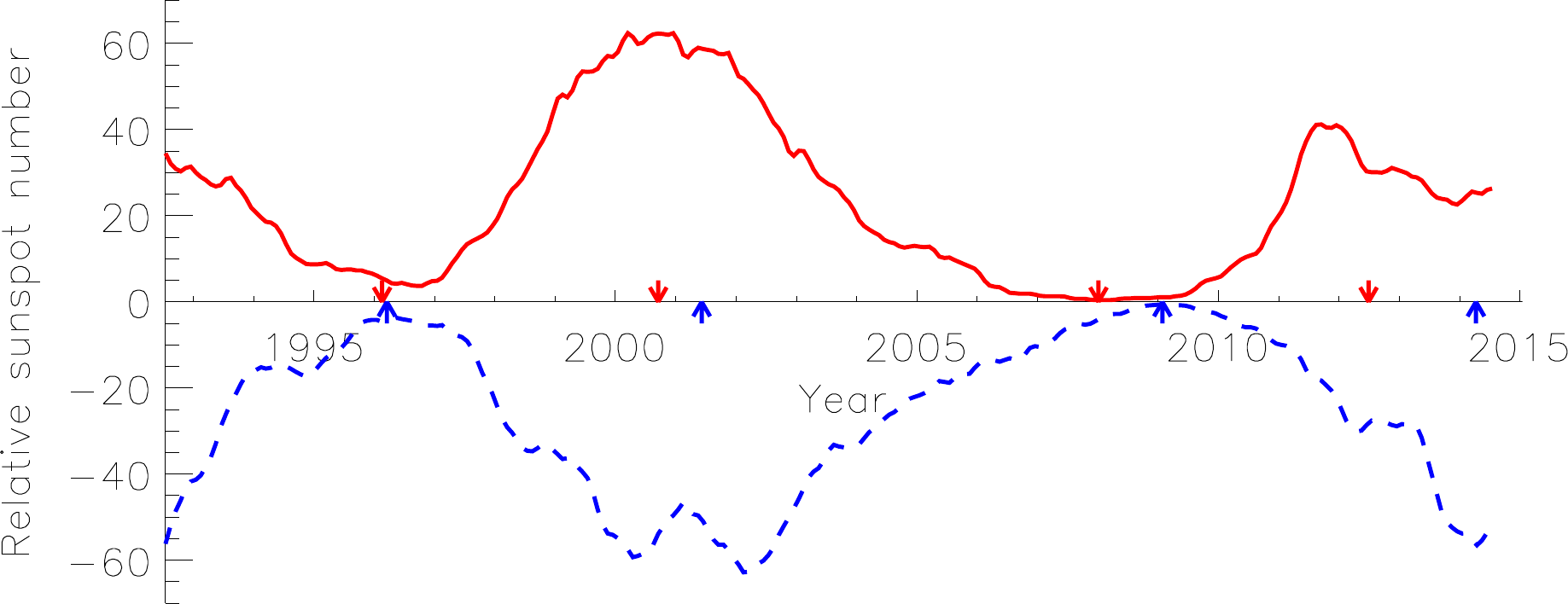}
\caption{Sunspot number in both hemispheres in cycles 23 and 24 (the sunspot number on the south was taken negative for illustration) with beginning, maximum, and the end of the cycles indicated. (From SILSO data, Royal Observatory of Belgium, Brussels)}
\label{fig:hemispheric_ssn}
\end{figure}
The pores in the intermediate distances (40--100~Mm, the \emph{intermediate pores} henceforth) from the closest SAR are especially interesting. These pores are those we were looking for. They have a correct sign to contribute to the reversal of the polar cap. They emerge in the regions of a weaker field, far enough from the neighbouring SARs. The total magnetic flux contained in the intermediate pores is of the order of $10^{21}$~Mx (\emph{on average} in the studied part of cycle 24, each Carrington rotation contains a total flux of $2.3\times 10^{21}$~Mx in the intermediate pores in the northern hemisphere and $-0.9\times 10^{21}$~Mx in the southern hemisphere). This number is roughly comparable to the total flux in the polar cap estimated to around $4\times 10^{21}$~Mx (see Section~\ref{sect:appearance}). A close investigation shows that the pores under discussion reside mostly on the trailing side of the closest SAR. 

It is not possible to perform an identical analysis for cycle 23 because it is covered only by MDI observations, which is a lower resolution instrument. The histogram of sizes of the pores shows that the vast majority of detected pores from HMI observations have sizes less than 16~px$_{\rm HMI}^2$, hence smaller than 1~px$_{\rm MDI}^2$. The number of pores detected in MDI data over 15 years of MDI observations is only 18\,683, while in the HMI data, the same routines detected 118\,540 pores over 4 years. This causes figures similar to Fig.~\ref{fig:polarity_match} constructed for MDI to appear very noisy with many gaps and generally not useful for any serious analysis. 

Despite the inability to assess the polarity trends during cycle 23, we may use the findings obtained using HMI observations and focus on intermediate pores only. We investigate the flux bias in those pores averaged over the rising and declining phases of cycle 23 and the rising phase of cycle 24. The splitting was done separately for each hemisphere by using the international smoothed hemispheric sunspot number by fitting a parabola to the points around the suspected minimum and maximum of the given cycle (see Fig.~\ref{fig:hemispheric_ssn}). 

\begin{figure}[h]
\centering
\includegraphics[width=0.49\textwidth]{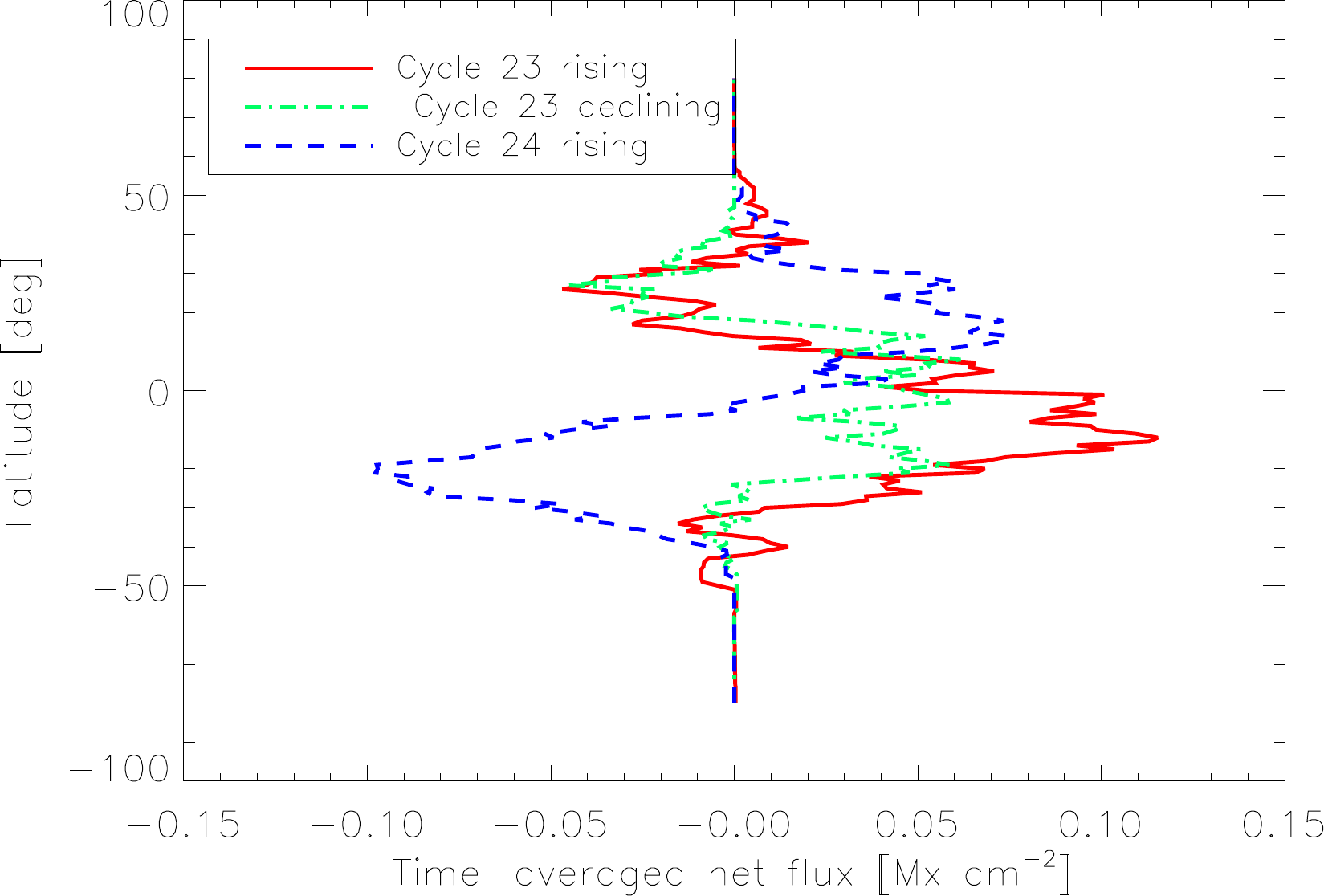}
\caption{Net magnetic flux in the pores in intermediate distances averaged over a declining or rising phase of cycle 23 and 24. }
\label{fig:hemispheric_flux_phases}
\end{figure}

\begin{table}[h]
\caption{Prevailing signs of the magnetic field of various features in the rising and declining phases of cycle 23 and rising phase of cycle 24 on the northern and southern hemisphere. }
\centering
\begin{tabular}{l|ccc}
\hline
Feature & $\nearrow$23  & $\searrow$23 & $\nearrow$24 \\
\hline
p-spot N & $+$ & $+$ & $-$ \\
pores N & $-$ & $-$ & $+$ \\
mid-lat N & $-$ & $-$ $\nearrow$ $+$ & $+$ \\
cap N & $+$ & $-$ & $-$ \\
\hline
p-spot S & $-$ & $-$ & $+$\\
pores S & $+$ & $+$ & $-$ \\
mid-lat S & $+$ & $+$ $\searrow$ $-$ & $-$ \\
cap S & $-$ & $+$ & $+$\\
\hline
\end{tabular}
\label{tab:polarities}
\end{table}

The time-averaged net magnetic flux in the intermediate pores in the three discussed phases are displayed in Fig.~\ref{fig:hemispheric_flux_phases}. It is evident that the signs of the net fluxes alternate with hemisphere. It is particularly interesting to see that the dominant polarity in the pores during the rising phases are opposite to the polarity of the polar caps and that the polarities then match during the declining phase. This shows the possible ability for those particular pores to reverse the polar magnetic field. We also find that the sign of the net magnetic flux in the mid-latitudes (obtained from the magnetic butterfly diagram by averaging it over the latitudinal bands between 30 and 50 degrees) corresponds to the polarity in those pores. The signs of the prevailing magnetic field in various magnetic features (leading spot in the active regions, pores in the intermediate distances from the nearest SAR, mid-latitudinal average magnetic flux, and the polar cap) in the rising and declining phases of the two investigated cycles are summarised in Table~\ref{tab:polarities}. Both Fig.~\ref{fig:hemispheric_flux_phases} and Table~\ref{tab:polarities} thus seem to indicate a possible role of those intermediate pores in the polar field reversal at the maximum of the cycle. Indeed, the flux contained in those pores seems to be carried towards the polar caps until this flux is strong enough to possibly participate in the polar field reversal. The magnetic field in the intermediate pores then continue to be amplified during the declining phase of the cycle before changing sign at the beginning of the rising phase of the next cycle.

\section{Link between pores and bipolar magnetic regions}
\label{sect:bmr}

The BL surface term relies on the presence of tilted bipolar magnetic regions, which may or may not contain sunspots. By considering only SARs in this study we ignore the presence of bipolar magnetic regions (BMRs) without sunspots and possibly introduce a selection effect in interpretation of our results. Isn't it that the intermediate pores all originate from smaller BMRs?

We wrote an additional code to detect BMRs in full-disc magnetograms. We could in principle use the outputs of the HMI Active Region Full-Disk Masks\footnote{http://jsoc.stanford.edu/jsocwiki/ARmaskDataSeries} pipeline and related products, however they do not cover the MDI era, which a significant portion of the minimum between cycles 23 and 24. We have however verified that our pipeline yields comparable results.

Our BMR detection code works as follows: To remove the noise, the magnetograms were first smoothed with a Gaussian window with full-width-at-half-maximum of 15~Mm. The code was based on segmentation of the compact patches of negative and positive polarities above the threshold and then pairing them together. We make different choices for the threshold value, the lowest chosen threshold is 10~Gauss, which is quite low so that we believe we do detect also small magnetic regions and our sample is thus almost complete down to less than the supergranular scales. As a BMR a pair of the positive and negative patches was marked, which had a smallest distance within the given full-disc magnetogram. The distance metric was modified by penalisation of the distance in the meridional direction, so that the segmented pairs stretching in the zonal directions were preferred. 

What is the difference between SAR and BMR? The SARs were obtained by constructing the masks based on the full-disc intensitygrams, whereas BMRs come from full-disc magnetograms. By construction, each SAR lies within some BMR, BMR is usually larger by a magnetised rim around SAR. Additionally, we have a large number of BMRs that do not coaling to any considered SAR, as they do not possess sunspots. 

\begin{figure}[h]
\centering
\includegraphics[width=0.49\textwidth]{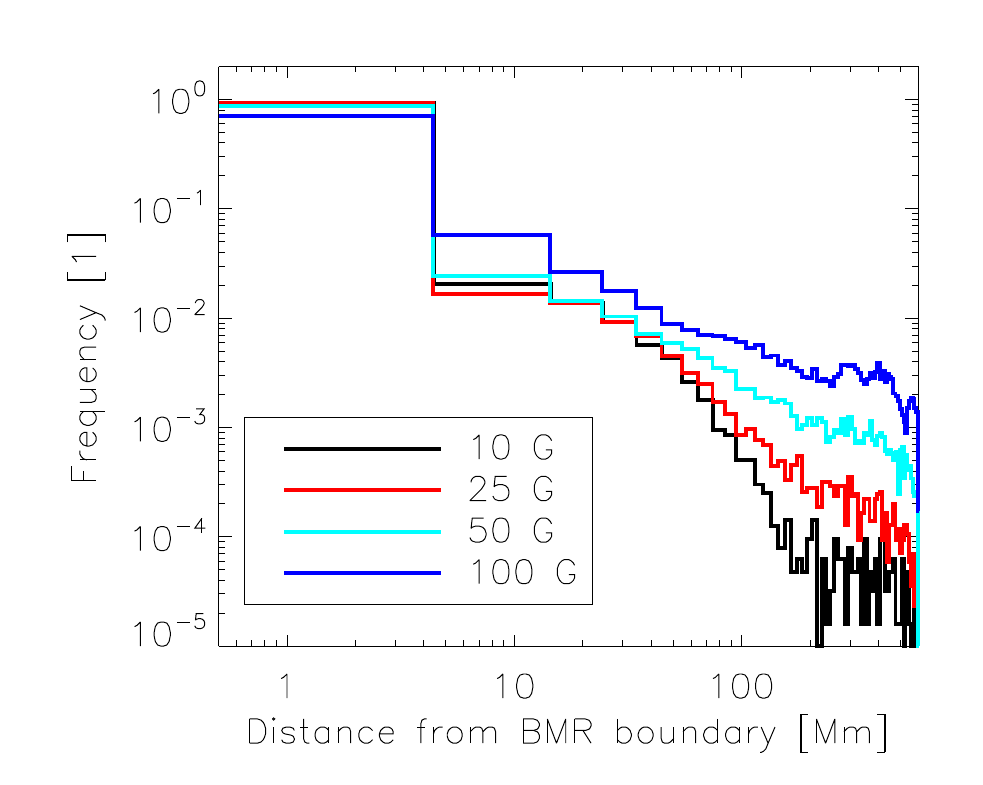}
\includegraphics[width=0.49\textwidth]{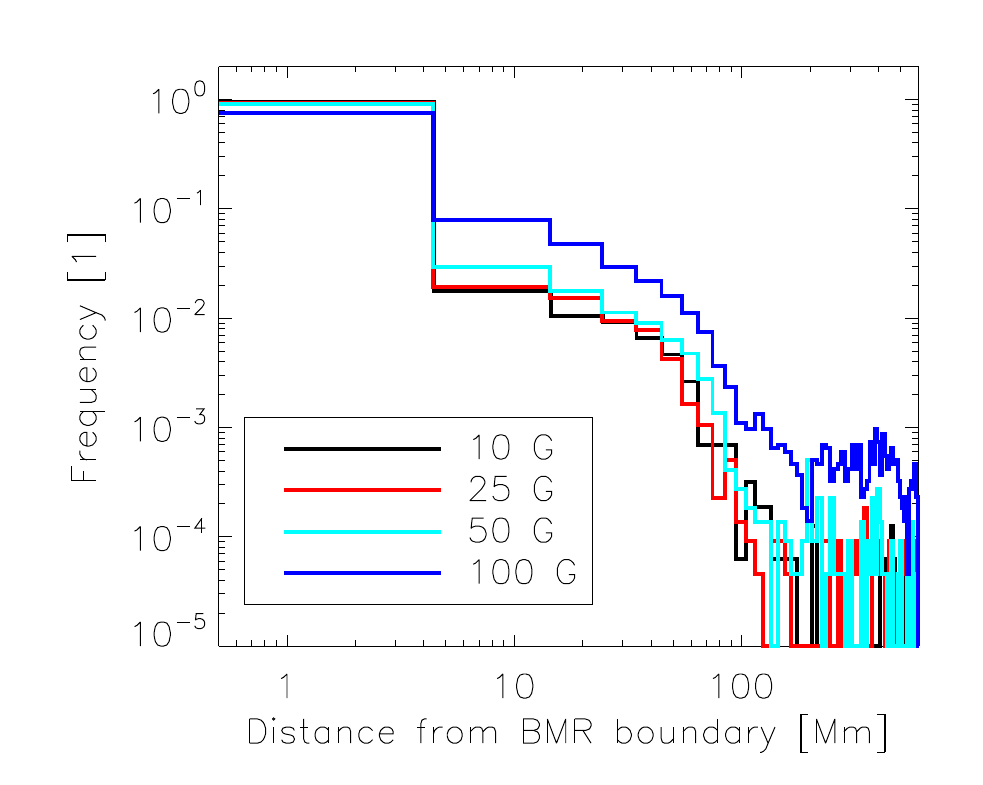}
\caption{Histogram of distances of the pores from the closest boundary of BMR for various thresholds in BMR detection. Upper panel -- all pores outside SARs, lower panel -- intermediate pores only. The peak at the distance 0~Mm represents pores located within bipolar magnetic regions detected by the automatic routine. Notice a logarithmic scales.}
\label{fig:pore_bmr_distances}
\end{figure}

For each pore considered in the previous analysis we measured its distance to the closest edge of any BMR. In case the pore was located within a BMR, this distance was set to zero. The distances were measured for several thresholds in the BMR detection code as described above. The histograms of these distances are displayed in Fig.~\ref{fig:pore_bmr_distances}. For the 10-G threshold, 92\% of the pores outside SARs are located within one of the BMRs, 6\% are closer than 30~Mm to the closest BMR and less than 1\% of the pores outside SARs lie farther than 60~Mm from the closest BMR. Analogical histogram constructed for the intermediate pores only is very similar with an even faster decay towards larger distances (Fig.~\ref{fig:pore_bmr_distances} right). 

\begin{figure*}[!t]
\centering
\includegraphics[width=0.7\textwidth]{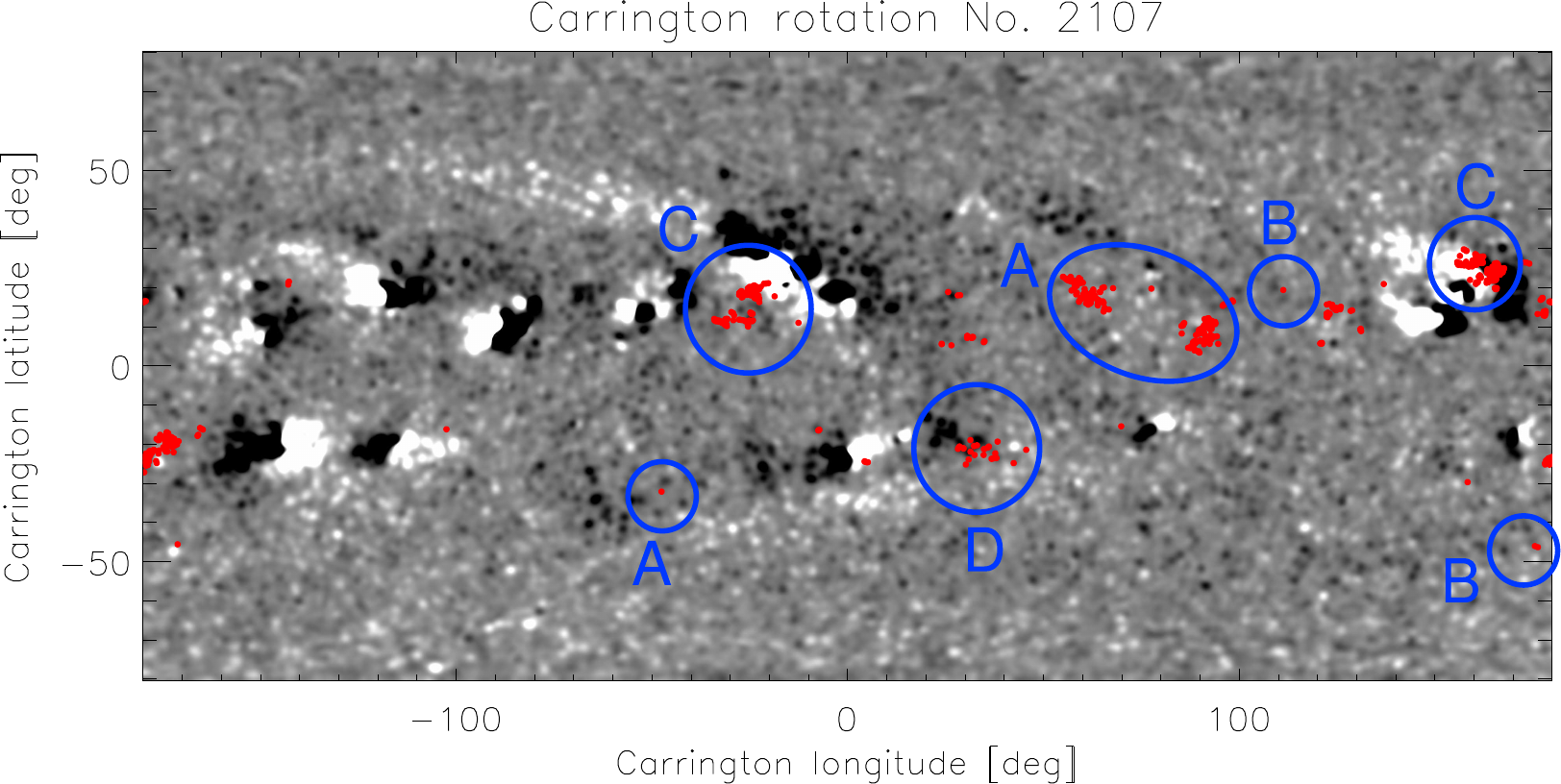}
\caption{An example synoptic map with various locations of intermediate pores with respect to the closest bipolar active regions. A -- the pores can be found in the streams of trailing polarity extending from the dispersed active regions towards poles. B -- pores located far from any obvious patch of organised polarity. C -- pores in or very close to the trailing polarity of bipolar magnetic regions (most of the pores can be found here). D -- Pores in or very close to the leading polarity of bipolar magnetic regions. }
\label{fig:CR2107}
\end{figure*}

This suggests that pores outside SARs are indeed a proxy for organised bipoles in the photosphere and therefore for a weak BL term. This conclusion seems quite insensitive to the choice of the threshold parameter in the BMR detection code. The decay towards larger distances from BMR is less steep for larger thresholds (when lesser number of BMR is detected), but even in this case 57\% of all pores outside SARs (64\% of intermediate pores) lie within one of the detected BMRs. 
\begin{figure*}[!t]
\centering
\includegraphics[width=0.9\textwidth]{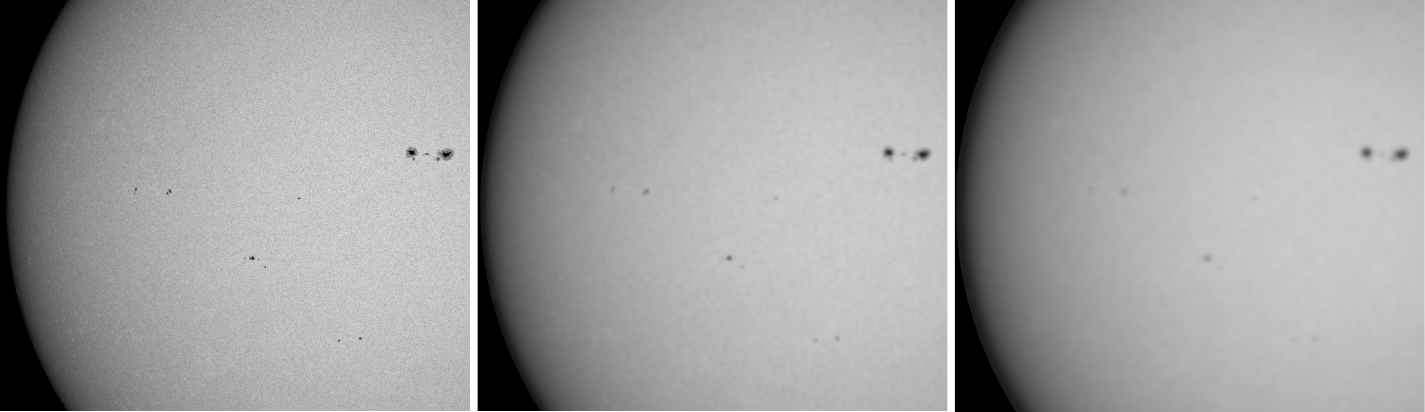}
\caption{What could ancient observers possibly see? Left -- contemporary white-light image of the Sun captured by HMI from space. Middle -- a simulation of what could observers see assuming 2" seeing and 2" resolution telescope. Right -- the same for 2" seeing and 5" telescope. Obviously, the large spots remain perfectly observable, however the small ones are blurred and may be easily missed by the observers, especially when observing at low-contrast conditions (e.g. in a free space).}
\label{fig:ancient_simulation}
\end{figure*}

A visual inspection suggests that the pores outside SARs may be found either in or next to spotless BMRs, or in the polarity streams migrating from the activity belt towards the poles (see Figure~\ref{fig:CR2107}). In the consequent Carrington rotation the pores typically appear at similar location, creating some sort of persistent nests of pores with a persistence time similar to the lifetime of the adjacent large-scale BMR. 

\section{Temporal link between bipolar magnetic regions and active regions with sunspots}
So far we considered only the spatial distance of the pores from SARs, however it is plausible that the pores we detect are located in the remnants of SARs, i.e. we should also consider a ``temporal distance''. In the previous Section we established that a vast majority of the pores under study originate from bipolar magnetic regions. Therefore we investigated statistically, whether these BMRs are strictly remnants of SARs or whether they can exist on their own. 

Already rough numbers indicate the latter option. For the studied period the automatic algorithm detected 30\,418 SARs, whereas a lot more BMRs (depending on the chosen threshold: 26\,803 for the 100~G threshold, 57\,247 for 50~G, 108\,087 for 25~G, and 207\,780 BMRs for the 10~G threshold) were detected. To obtain qualitative results, we investigated the temporal and spatial coalignment of each BMR with all SARs. 

For each BMR and also for each SAR we computed heliographic coordinates and also other descriptive quantities (such as the unsigned total magnetic flux in the BMR). Then we looped over a complete set of BMRs (detected for the 10~G threshold) and searched for a SAR, which was closest to the given BMR's location any time in the past 120~days (more than four solar rotations). The considered distance metric was a distance of gravity centres of the BMR and SARs. By this approach we investigate the possibility of the BMRs being the remnants of SARs.

\begin{figure}[!ht]
\includegraphics[width=0.49\textwidth]{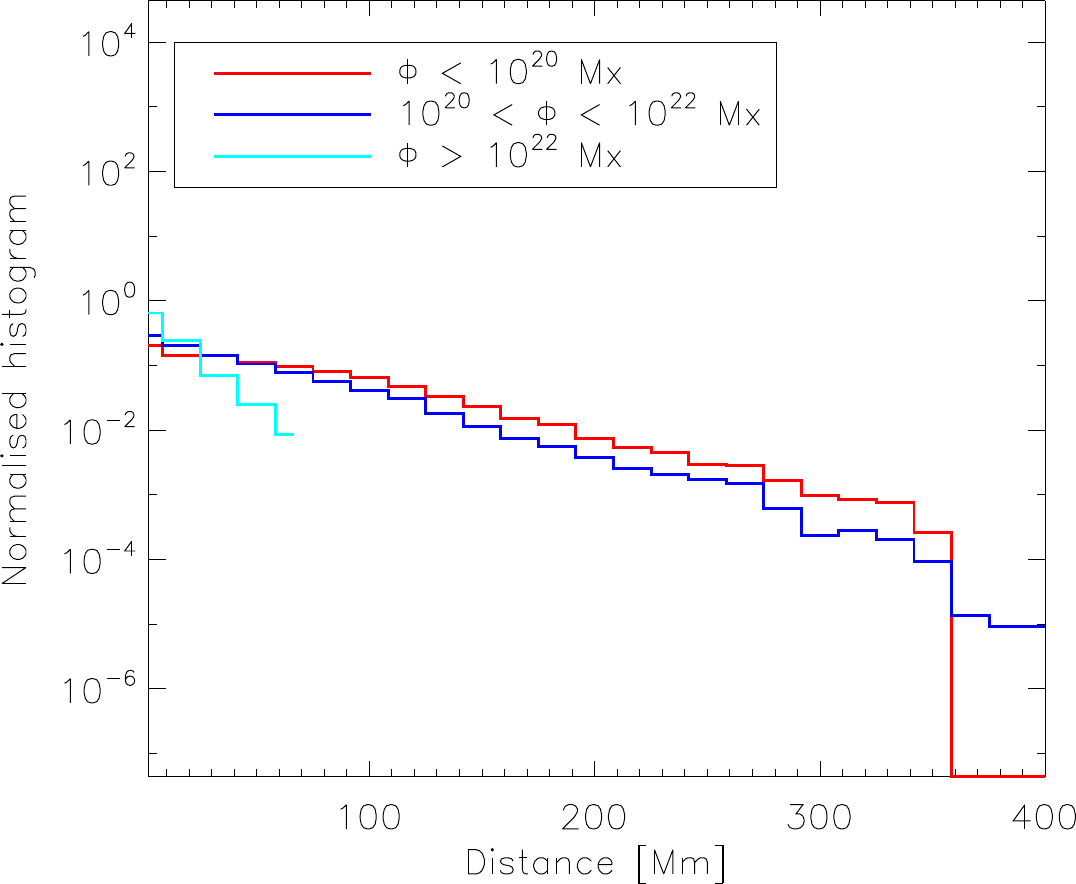}
\caption{Histogram of distances between BMRs and past SARs for three classes of BMRs distinguished by the total unsigned magnetic flux.}
\label{fig:bmr_vs_sar}
\end{figure}
The results are presented in a form of histogram of distances between the BMR and the closest SAR in the past in Fig.~\ref{fig:bmr_vs_sar}, which is derived for different classes of BMRs distinguished by the total unsigned flux $\Phi$. Should a vast majority of BMRs be remnants of SARs, the histogram of distances would be strongly peaked around zero. We see that generally, this is not the case. The conclusions depend on the total magnetic flux in the given BMR. BMRs with a large total unsigned flux is large, larger than $10^{22}$~Mx, are almost solely remnants of SARs. In our sample, their centres of mass are located less than 70~Mm away from a centre of mass of a previously existing SAR, which existed at this location within the past 120~days. On the other side of the spectrum, BMRs with a small total unsigned magnetic flux, smaller than $10^{20}$~Mx, have a large distribution in distances, basically saying that they can form on their own and are not linked to classical active regions with sunspots. As seen e.g. in Fig.~\ref{fig:CR2107} (feature B), such regions also contain pores. These BMRs may be considered as ``failed emergence'' \citep[e.g.]{1965ApJ...141.1492B,1965ApJ...141.1502B} of a raising $\Omega$ loop, in which proper sunspots could not form but in which pores can exist.

\section{Discussion}
\label{sect:conclu}

Many geomagnetic indices indicate that organised magnetic field must have existed in the solar photosphere even during the Maunder minimum, however there is an obvious lack of positive sunspots observations. The resolving power of instruments of observers of the Maunder minimum era may provide a hint towards understanding the recorded lower activity during this period. Given the expected resolving power of the instruments used in the 17th century of some 2"--5" we may expect, magnetic features such as pores or weak bipolar magnetic regions could have been missed by the observers (for an idea see Fig.~\ref{fig:ancient_simulation}). This estimate of the resolution limit is supported by a recent thorough investigation through the archival observations of Gustav Sp\"orer, who observed the Sun regularly in 19th century, hence two centuries after the Maunder minimum. The authors show that his observing limit was around 4" \citep{2014arXiv1411.7790D}.

Indeed when studying the archives, Hoyt \& Schatten described the situation several times in their articles, when a sunspot was observed by some astronomers on a particular day, but not by others on the same day. If we neglect the possibility of an observer's error, it may be that those spots were short-lived, so that they appeared and disappeared during the same day. Similarly, some of these spots could have been small, so that the telescopes of those days resolved those spots for some observers and not for others. Both quantities again resemble the usual properties of solar pores. 

A further independent indication that such transition from large to small spots might be operating on the Sun was published by \cite{2012ApJ...758L..20N}. The authors found that as the overall strength of the solar cycle decreases secularly \citep{2006ApJ...649L..45P}, there is a transition in the frequency of occurrence of the spots from large ones becoming rare to small ones becoming more common. This conclusion is supported also by a consequent study by \cite{2014ApJ...794L...2K}.  

In order to understand how the lack of sunspot detection is compatible with the  continuing existence of a cyclic 11-yr global magnetic field reversal, we have investigated the properties of pores emerging (or forming) at the solar surface during a magnetic cycle, focusing on the ability of such pores, after their decay, to participate to the net magnetic flux needed to reverse the polar caps. Such pores, as explained above, could have been missed due to a lack of resolution power of the instruments used during the Maunder minimum era. We have thus compared the flux carried by the pores outside SARs to the one contained in polar caps as well as studied any polarity trend of those structures. We indeed found that pores within a distance of 40 to 100~Mm (possibly to 140~Mm) from the closest SAR do possess the required polarity bias and that most of them are located on the trailing side of the nearest active region. The similarity of the shape of their butterfly diagram and of the butterfly diagram of sunspots also suggest that they could contribute to the operation of the solar dynamo. We also note that in our study no high latitude pores were detected beyond the limit of the activity belt.

We further find that the pores with the correct polarity trend in the rising and declining phase of solar cycle 24 are often found inside weak bipolar magnetic regions (see Fig. \ref{fig:pore_bmr_distances}). Such weak BMRs, in the sense that no sunspots were able to form within them, certainly contribute to the polarity trend found in the intermediate distance pores. Possibly the fragmentation of weak flux tubes (mostly of their trailing leg) forming these weak BMR's after their emergence may be responsible for the occurrence of the pores under study and of their polarity trend. A giant-cell convection may be responsible for such fragmentation and hence our finding that the pores with a polarity bias are predominantly located 40--100~Mm (possibly 40--140~Mm) from the closest large active region with sunspots may provide a hint on a length scale of such mode of convection. A ``Magnetic range of influence'' of emergent flux was investigated by \cite{2014ApJ...784L..32M} with a conclusion that the length scale of such range of influence is between 100--200~Mm. Also other claims for detection of cellular features of a similar length scale may be found in literature. 

The link between surface magnetic field and the internal dynamo that produces it is complex and many scenario to explain this link have been proposed over the years \citep{2010LRSP....7....3C, 2012LRSP....9....6M}. One scenario has attracted much attention in the last 20 years, the so-called flux-transport model that relies on the Babcock-Leighton mechanism \citep{1961ApJ...133..572B,leighton69}. Such solar model rests on the observations that much of the magnetic flux is advected by the meridional flow and/or diffused towards the poles. This results first in a cancellation of the polar cap magnetic flux in the rising phase of the cycle and then to the strengthening of the new polarity cap in the declining phase after the global field reversal has occurred \citep{2004AA...428L...5B, 2012ApJ...753..157S}. Near the equator the trans-equatorial flux cancellation of opposite field polarity helps renewing the global solar magnetic field polarity \citep{2015ApJ...808L..28J}. More specifically two types of models have been developed over the years that use magnetic flux transport mechanisms as a main ingredient to explain the solar magnetic field and the 11yr cycle: surface $(\theta,\phi)$ flux models \citep{1990ApJ...365..372W, 2002ApJ...577.1006S, 2012LRSP....9....6M, 2014ApJ...791....5J} and meridional $(r,\theta)$ mean field dynamo models \citep{1999ApJ...518..508D, 2007AA...474..239J, 2011Natur.471...80N, 2014SSRv..186..561K}. New efforts that attempt to couple the two approaches are being undertaken \citep{2014ApJ...785L...8M}. In such flux transport scenario the filling factor, size, amplitude and life time distribution of the magnetic features is essential in order to understand the solar surface magnetism. It would be interesting to adapt our finding of a systematic polarity trend for pore-like structure to see how efficient it is at reversing weak polar caps.

Of course one must be careful when extrapolating our contemporary study of the pores' distribution and polarity trends back into the Maunder minimum era. We do not attempt here to make a direct comparison but to understand what the poorer resolution of the observations implied back then in terms of missing key surface magnetic features. We conclude that it could be the case that during the Maunder minimum the Sun could have weak BMR’s or small pores that were unobserved but still played a role since we know \citep[thanks to $^{10}$Be content in ice cores][]{1998SoPh..181..237B} that the 11yr cycle dynamo was operating.

So, even though the current solar magnetic state is unlikely representative of a minimal state of the Sun, we get some confidence in our analysis that the Sun is not in a maximum state of activity by noting a) the abnormal length of the transition between cycle 23 and 24 b) the associated high number of spotless days (800) and c) that cycle 24 is much weaker than the last 5 cycles \citep{2012JSWSC...2A..06C}. Further there are some claims that the Sun is going to end the current Gleissberg cycle by entering a quieter state with less spots  \citep{2008GeoRL..3520109A, 2012ApJ...757L...8L}. Since the threshold of 1500 G is necessary for the appearance of dark area (umbra) \citep{1970SoPh...13...85S, 2012ApJ...757L...8L} a weaker state of activity is likely to lead to a change in the distribution of magnetic features.

\section{Conclusions}
We thus suggest that emerging magnetic field during the grand minimum did form weak BMRs and associated pores that were undetected. We find that intermediate distance pores (between 40--100 Mm from the closest SARs) are good proxies to assess the polarity trends akin to a weak Babcock-Leighton source term needed to contribute to the polar cap field reversal. The reason why such pores and weak BMR's were undetected is likely due to a lack of resolving power and to the fact that the emerging surface magnetic field was not strong enough to form large stable sunspots, perhaps except for a few cases. We show in this work that the appearance of pores outside SARs follows the solar cycle in terms e.g. of the position on the magnetic butterfly diagram. We also find that the intermediate pores with the correct polarity trend further possess an amount of flux compatible with the flux in polar caps during weak cycles, even it is likely that only a fraction of the flux found in pores will be carried poleward. On the other hand, these pores will certainly be surrounded by a larger patch of the magnetic field, which will partly be also carried polewards. We conclude from our pores study that a process akin to a weak Babcock-Leighton magnetic source term could have contributed to the operation of the solar global dynamo during the Maunder Minimum.

\begin{acknowledgements}
M.\v{S}. acknowledges the support of the Czech Science Foundation (grant 14-04338S) and of the institute research project RVO:67985815 to Astronomical Institute of Czech Academy of Sciences. A.S.B. acknowledges financial support by CNES Solar Orbiter grant, CNRS/INSU Programme National Soleil-Terre, ERC STARS2 207430 and SolarPredict 640997 grants and wishes to thank M. DeRosa for useful discussions.
\end{acknowledgements}

\bibliographystyle{aa}
\bibliography{biblio}

\appendix
\section{Special data treatment}
\label{sect:appendixa}
As we mentioned above, not only the real solar pores are detected by the automatic detection routine, but also artifacts such as the dust or perhaps even bad pixels. These artifacts look exactly like the pores and thus are detected by the algorithm as false positives. That is true especially for MDI in last years of its operation. They are removed using a simple assumption that their positions on the CCD do not change with time. Hence all the pores are looped over and when there is a clustering at the same positions, all such representatives are removed from the set. Such correction removes a lot of false positives from the MDI-data-based detections of the pores (in the last years of operation it even is a vast majority of detections), but has a negligible effect to HMI-data-based detections of pores. 

MDI and HMI are different resolution and sensitivity measurements. Two measures are taken to mitigate with this issue:
\begin{enumerate}
\item Only features having comparable linear sizes (in Mm) are further used for analysis. The size of MDI pixel is around four times larger than the size of HMI pixel. Hence two thresholds are taken for pores analysed further: a minimum area of the pore must be 1~px$^2$ for MDI and 16~px$^2$ for HMI, maximum area of the considered pore must be below 7.5~px$^2$ for MDI and 120~px$^2$ for HMI. This way the different resolutions are dealt with. 
\item The different sensitivity in line-of-sight magnetic field measurement is dealt with using a calibration from the overlap period (24 April 2010 to 10 April 2011). Only observations performed by both instruments at the exactly same time (an allowed deviation is 5 minutes) are considered and also pores detected by both instruments were taken into account. These are strong constraints, as only 202 pores fulfilled them. The considered pores are evenly distributed over the longitudes but are more concentrated to the northern hemisphere. Then the average magnetic field intensity in the detected pores is compared from the two instruments. There is a large correlation between the two (Pearson's correlation coefficient 0.83) and the slope of the linear fit MDI to HMI data is 1.49 (the fit was performed by assuming that both measurements have a random nature). Hence in a further analysis, the magnetic fluxes determined from HMI are divided by this number.When the pores of negative and positive polarities are treated separately, smaller correlation coefficients ($\sim0.3$) are found and the slope of the fit is also smaller. Therefore the absolute numbers presented in the paper as the values for the total flux etc. may be by a factor of the order of unity different based on the calibration slope taken. This factor however does not change the conclusions. We note that if the factor were say twice smaller than the used value, it would introduce a visible vertical stripe in the magnetic butterfly diagram of pores (Fig.~\ref{fig:mgbtfly}) when the transition from the MDI to HMI data occurred.
\end{enumerate}
An additional constraint to deal with badly assessed detections is that the heliocentric angle must not be larger than 80 degrees. There are often artifacts detected as pores at the limb, which are removed by this constraint. 

In total, 49\,724 individual observations were analysed, 35\,040 of this from HMI. 

\section{Effect of the sampling in time}
We sample the observations with 6-hr sampling. What if we miss some of the pores due to their limited lifetime? What if the pores farther from the equator live shorter than those in the equatorial region? In that case we would introduce a systematic error in the pores' distribution at various distances from the equator and hence for example miss most of the pores at high latitudes. Currently, we do not observe pores there. 
 
To investigate the effect of the sampling interval, we performed a simple test: the HMI data covering months May 2010 and January 2012 were analysed twice using the same routines but with different time-steps between the consecutive observations (1 hour and 3 minutes). The aim is to search for systematic differences in the distribution of pores in latitudes. This procedure has lead to the following results:

\begin{itemize}
\item The distribution of pores' locations in latitudes does not change. Shorter time-step does not allow to detect pores at high latitudes. So either there are no pores or the detection routine can't detect them. The latter reason seems probable. Due to the projection effect, the width of the 2-Mm pore will decrease below the spatial resolution of HMI at already some $45\degr${} from the disc centre. 
\item The obtained flux maps are highly correlated ($\rho \sim 0.9$), so the distribution of the flux over the photosphere of the Sun does not depend on the time-step
\item The total magnetic flux is 50\% larger for time-step of 3 minutes than that for the time-step of 1 hour. This would suggest that there occurs a lot of flux cancellation within 1 hour. 
\item The total number of detected pores is expected to be 20-times larger from the 3-min-sampled data than from the 1-hr-sampled data (there is 20 non-overlapping 3-minute intervals in one hour). The test showed that the multiplier is exactly 20 in May 2010. In January 2012 it gets a value of 18.7, hence there is less pores detected from the 3-min data than expected. This could be explained by the formation of the pores, when the contrast of a forming or decaying pore with respect to the quiet-Sun background is lower than our threshold. 
\end{itemize}
We conclude that the time-step has only a weak effect. The non-detection of the pores at higher latitudes is probably due to the projection effect in combination with the spatial resolution of the instruments. 

\end{document}